# Design of Microresonators to Minimize Thermal Noise Below the Standard Quantum Limit

S. Sharifi[1,2,3,4], Y. M. Banadaki[4,5], T. Cullen[1], G. Veronis[2,3], J. P. Dowling[1,4], and T. Corbitt[1,4]

[1]Department of Physics and Astronomy, Louisiana State University, Baton Rouge, Louisiana 70803, USA
[2]School of Electrical Engineering & Computer Science, Louisiana State University, Baton Rouge, Louisiana 70803, USA
[3]Center for Computation and Technology, Louisiana State University, Baton Rouge, Louisiana 70808, USA
[4]Hearne Institute for Theoretical Physics, Louisiana State University, Baton Rouge, Louisiana 70803, USA
[5]Department of Computer Science, Southern University, Baton Rouge, Louisiana 70813, USA

**Abstract**

We present a design for a new microresonator whose geometry is optimized to maximize sub-Standard Quantum Limit (SQL) performance. The new design is predicted to have thermal noise well below the SQL across a broad range of frequencies when operated at 10K. The performance of this designed microresonator will allow it to serve as a test-bed for quantum non-demolition measurements, and to open new regimes of precision measurement that are relevant for many practical sensing applications, including advanced gravitational wave detectors.

## 1. Introduction

One hundred years after Albert Einstein predicted the existence of gravitational waves in his general theory of relativity [1], the Laser Interferometer Gravitational-Wave Observatory (LIGO) [2] and advanced Virgo [3] have opened a new window to the universe by detecting the first gravitational waves in 2015 [4]. These discoveries impel the need to develop strategies for improving the rate of detections by decreasing the limiting noise sources in gravitational wave interferometers [5-8]. Quantum mechanics dictates that the precision of physical measurements is subjected to certain constraints. The Standard Quantum Limit (SQL) sets a limit for conventional continuous interferometric displacement measurements. The SQL balances the imprecision of the measurement from photon statistics with the unwanted quantum back action (radiation pressure noise). Pushing the displacement measurements towards this limit is crucial for improving systems ranging from nanomechanical devices [9], ultracold atoms [10], and Advanced LIGO [11]. To go beyond the SQL, correlations between shot noise and radiation pressure noise must be created. Various techniques have been proposed to achieve this, including speed-meter interferometers [12], which monitor the relative momentum of the test-mass mirrors, and using optical springs [13], which manifest as a restoring force created by the optical field. Recently, Mason *et. al* [14] exploited strong quantum correlations in an ultra-coherent optomechanical system and demonstrated off-resonant force and displacement sensitivity reaching 1.5 dB below the SQL.

Quantum noise, including quantum radiation pressure noise (QRPN) [15], is one of the main limiting phenomena across a wide range of frequencies that originates from the quantum nature of photons in an interferometer. However, classical noise sources such as environmental vibrations and thermally-driven fluctuations are typically large and can prevent sub-SQL measurements. Approaching sub-SQL measurements in optomechanical systems has relied on techniques [16-18] aiming to first reduce the thermal motion and then test the proposals for reducing QRPN [19-25]. The non-negligible thermal noise (TN), which is governed by the fluctuation-dissipation theorem [26], sets a limit on the motion of mechanical resonators [27].

In a recent paper [28], we presented a broadband and off-resonance measurement of QRPN in the audio frequency band. We developed low-loss single-crystal microresonators with sufficiently minimized thermal noise so that the QRPN was observed at room temperature. The noise spectrum obtained showed effects due to QRPN between about 2 kHz and 100 kHz. In this paper, we use a micro-genetic algorithm and a finite element method (FEM) based model to design new GaAs/AlGaAs microresonators, to study the effects of different geometries of the mirror pad and cantilever microresonator, and to minimize TN well below the SQL. The new microresonators will enable broadband, off-resonance sub-SQL experiments and serve as a test-bed for sub-SQL and back action evasion techniques. In addition, this optomechanical device can operate as a filter cavity inside the signal-recycling cavity in the unstable regime [29]. The unstable filter acts as a phase compensator with negative dispersion in the single loop where a feedback control stabilizes the entire system. Within the loop, free propagation inside the arm cavity causes phase lag of the sidebands that can be compensated by the unstable filter. Minimizing the thermal noise of the unstable microresonator filters enables the implementation of this scheme and enhances the bandwidth of advanced gravitational wave detectors.

This paper is organized as follows: In Sec. II we describe the theory of designing the new microresonators using the micro-genetic algorithm and the model equations of TN and SQL that are used to evaluate the measurement of the new microresonators at the sub-SQL regime. This section is followed by a discussion of the effect of changing the geometry of the mirror pad and microresonator cantilever on the sub-SQL region of displacement measurement. Finally, our conclusions are summarized in Sec. IV.

## 2. Design of New Microresonators

Figure 1 shows the optomechanical system consisting of a Fabry–Pérot cavity with a mechanical oscillator as one of the cavity mirrors. The optomechanical cavity has a high-reflectivity, single-crystal microresonator that serves as the input coupler and a half-inch diameter mirror as the back reflector. A Nd:YAG laser beam is used to both stabilize the optomechanical cavity and measure the mechanical motion of the microresonator. The cavity is locked using a feedback loop that utilizes the restoring force produced by a strong optical spring [30]. The details of the optomechanical system can be found in Ref. [30]. Here, we focus on designing a new microresonator and on studying the effect of varying its geometry on the sub-SQL region. While the mirrors currently used in gravitational wave detectors are based on silica/tantala coatings [31, 32], GaAs/AlGaAs crystalline coatings experimentally show great promise for reducing Brownian coating thermal noise [33-35], and are therefore being considered as coatings for the next-generation gravitational wave detectors.

Figures 2(a) and 2(b) show schematics of the cross-sections of the microresonator mirror pad structures that are designed from alternating layers of GaAs and $Al_{0.92}Ga_{0.08}$ as in Ref. [28], and as in this paper, respectively. A multilayer stack of alternate GaAs and $Al_{0.92}Ga_{0.08}As$ thin films used as high and low index materials forms a Fabry–Pérot cavity or Bragg mirrors [36]. In the new microresonator proposed in this paper, the thickness of each GaAs/$Al_{0.92}Ga_{0.08}$ quarter-wave optical pair in the Bragg mirror is kept the same as in Ref. 27, while the number of pairs is decreased to 22. The thick GaAs layer deposited over the InGaP buffer layer is removed and the last two layers are optimized by our recently developed hybrid optimization algorithm [37] consisting of a micro-genetic global optimization algorithm coupled to a local optimization algorithm [38]. Our goal is to obtain structures with thinner GaAs/$Al_{0.92}Ga_{0.08}As$ pairs while the transmission of 250 parts per million (p.p.m.) is kept same as in Ref. 27. Given the optical loss that we have observed in our optomechanical system, larger transmission will allow us to have a more pure measurement and consequently a more effective QRPN reduction technique, such as using squeezed light, that is extremely sensitive to optical loss [39]. The GaAs/ $Al_{0.92}Ga_{0.08}As$ pair under the InGaP layer not only contributes to the total reflectance of the mirror pad but also supports its weight in the cantilever microresonator. During the optimization with the genetic algorithm, we use a minimum thickness constraint for the thin GaAs/AlGaAs pairs, because too thin support layers may lead to a structure that is too fragile. The minimum thickness of

GaAs/AlGaAs pairs for the genetic algorithm is fixed to 50 nm based on the mechanical analysis of the microresonator geometries studied in this paper. Using the genetic algorithm for the proposed microresonator in Fig. 2(b) we found three GaAs/AlGaAs pairs with transmission of 250 p.p.m.: (1) $t_{GaAs}$ = 50.7 nm, $t_{AlGaAs}$ = 52.1 nm, (2) $t_{GaAs}$ = 41.2 nm, $t_{AlGaAs}$ = 40.9 nm, and (3) $t_{GaAs}$ = 35.8 nm, $t_{AlGaAs}$ = 34.7 nm. Figure 2(c) shows reflectance of 99.975% corresponding to transmission of 250 p.p.m. at the center wavelength of 1078 nm at which the structure was optimized for the previously proposed microresonator [27] and for the microresonator proposed in this paper with the thinnest GaAs/AlGaAs pair. The mirror center wavelength is red-shifted to 1078 nm to consider thermorefractive effects upon cooling and thereby the ultimate goal of operating these structures at cryogenic (liquid 4He) temperatures [40]. It can be seen that the new microresonator has the same bandwidth as the previous one, while the reflectance at higher and lower frequencies is smaller compared to the previous design.

We model the thermal noise using a FEM model of the microresonator in COMSOL. The FEM model of the cantilever uses the cantilever and mirror geometry as well as the mechanical properties of the GaAs/AlGaAs cantilever and mirror pad to compute the modal resonant frequencies. The mechanical modes that are most relevant for the work presented here are the first (fundamental), third (pitch), and fourth modes. We previously demonstrated [42] that the second (yaw) mode is not coupled to our measurement for the beam centered along the microresonator. Structural damping models contain a frequency-independent loss angle φ and for a harmonic oscillator have a displacement amplitude spectral density as follows [43]:

$$x(\omega) = \sqrt{\frac{4k_B T \omega_m^2}{Q_k m \omega \left[ (\omega_m^2 - \omega^2)^2 + \frac{\omega_m^4}{Q_k} \right]}}, \tag{1}$$

where $k_B$ is the Boltzmann constant, T is the temperature, m is the mass, Q = 1/φ is the quality factor, ω = 2π × f, and $\omega_m$ is the angular frequency of the mechanical mode. For structural damping, the thermal noise falls off as $1/\omega^{5/2}$ above the mechanical resonance frequency. Viscous damping, on the other hand, is proportional to $1/\omega^2$ above the mechanical resonance [43]. The SQL for the position measurement of a free mass with mass m is calculated based on the mechanical susceptibility $\chi$ using $x_{SQL} = \sqrt{\hbar |\chi|/2}$ [44].

## 3. Result and Discussion

Figure 3 shows the spectrum of TN and SQL that is calculated by summing the contribution of each mechanical mode in quadrature for the microresonators in Figs. 2(a) and 2(b) with different geometries at 10K (the lowest temperature possible to achieve for our equipment). All three microresonators have fundamental resonance frequencies between 100Hz and 1kHz, which are in the proper range for designing microresonators with improved sensitivities [28]. Compared to the 55 µm long microresonators (red and blue), the 97 µm long microresonator (black) not only has a larger number of higher-order modes but also the resonances are in the middle of the sub-SQL regime, which is not desirable. However, the short microresonators have a lower fundamental mode frequency due to their thin holders of the mirror pad (t = 70.5 nm).

The SQL/TN ratio as a function of frequency for the sub-SQL region of displacement noise is also shown in Fig. 3. Reducing the thermal motion at frequencies away from the resonance enables the sub-SQL regime and serves as a test-bed for very sensitive interferometric experiments. In order to study the sub-SQL region, we define four parameters: (1) the maximum ratio, $R_{MAX}$, (2) the frequency of maximum ratio, $f_{MAX}$, (3) the lowest frequency of the sub-SQL region, $f_L$, at which the TN is equal to the SQL, and (4) the bandwidth of TN and SQL equality (BWE) as $\log(f_H) - \log(f_L)$. Larger $R_{MAX}$ indicates a more secure range of sub-SQL regime and thereby better access to test and suppress the QRPN. Expanding the sub-SQL region(e.g. $f_{MAX}$ and $f_L$) to lower frequencies decreases heat generation from the laser beam in the cryogenic system and thereby facilitates the experimental setup. Broadening the sub-SQL range to higher frequencies is promising for future designs of an unstable filter cavity that operates beyond the frequency limit of 200 kHz in our current experimental setup. We study the effects of changing the geometry of the mirror pad and microresonator cantilever on the sub-SQL region for the proposed microresonator in Fig. 2(b).

### 3.1. Length of Microresonator Cantilever

Figure 4 shows the effect of varying the length of the microresonator cantilever. The frequency of the fundamental, pitch, and fourth modes as a function of the microresonator cantilever length for the three optimized microresonators is shown in Fig. 4(a). The resonance frequencies of the microresonators decrease as the length of the microresonator cantilever increases. In addition, the fundamental and pitch mode frequencies decrease more rapidly than the one of the fourth mode as the length increases. We observe that the thinner microresonator with $t_{GaAs/AlGaAs} = 70.5$ nm has the lowest frequencies for the fundamental, pitch, and fourth modes for any microresonator cantilever length. However, the dependence of mode frequencies to the thickness of GaAs/AlGaAs pairs increases for longer microresonators. The benefit of using a long cantilever is having a lower fundamental mode frequency and thereby lower thermal noise at the frequencies of interest. The two longest microresonators in our study with 55 µm and 100 µm length have fundamental resonance frequency between 100 Hz and 1 kHz which, as mentioned above, is the proper range for microresonators with improved sensitivities [28]. On the other hand, the microresonator with 100 µm length has more higher-order modes (e.g. the fourth mode) with frequencies below 200 kHz which is not a desirable feature [28].

$R_{MAX}$ increases from ~1.3 to ~ 4.6 and $f_{MAX}$ decreases from ~ 700 kHz to ~ 50 kHz as the length of the microresonator cantilever increases from 25 µm to 100 µm [Fig. 4(b)]. For length of 55 µm, the thinner microresonator with 70.5 nm thickness of GaAs/AlGaAs has the benefit of both larger $R_{MAX}$ and smaller $f_{MAX}$, and is thereby a promising design for a secure range of sub-SQL regime and less heat generation from the laser beam (i.e. broad frequency range shifted to lower frequencies and the displacement noise is well below that of SQL). The lowest frequency, $f_L$, decreases from ~ 450 kHz to ~ 10 kHz by increasing the length of the microresonator cantilever from 25 µm to 100 µm [Fig. 4(c)]. However, the lowest frequency $f_L$ steeply decreases to ~40 kHz as the length decreases to 55 µm, while it only slightly decreases as the length of the microresonator cantilever further increases [Fig. 4(c)]. In other words, $f_L$ is less sensitive to length variations for lengths longer than 55 µm. The BWE increases from 3.4 to 4.3 by increasing the microresonators' cantilever length from 25 µm to 55 µm [Fig. 4(c)]. However, the BWE decreases to 3.9 for length of 100 µm because the frequency range of SQL/TN>1 decreases due to the shift of higher-order modes to lower frequencies for longer microresonators. Similar to $f_L$, BWE is less sensitive to length variations for longer microresonator lengths for all three optimized microresonators [Fig. 4(c)]. Thus, for fixed parameters of w = 8 µm, r = 32 µm, y = 3.75 µm in Fig.1, we need to choose the longest microresonator (l = 100 µm) to have more secure sub-SQL range with the desirable values of low resonance mode frequencies, high $R_{MAX}$, low $f_{MAX}$, low $f_L$, and relatively high BWE.

### 3.2. Width of Microresonator Cantilever

Figure 5 shows the effect of varying the width of the microresonator cantilever. Decreasing the width of the microresonator cantilever from 12 µm to 6 µm lowers the frequency of the fundamental and pitch modes by ~30% as shown in Fig. 5(a). The frequency of the fourth mode has an insignificant dependence on the width of the microresonator cantilever. The shift of the modes to lower frequencies as the cantilever width is varied is not as significant as the shift caused by changing the length of the cantilever. All three microresonators satisfy the constraint of having the fundamental resonance frequency between 100 Hz and 1 kHz, as well as the higher-order modes above 200 kHz for all cantilever widths considered [28]. The thinner microresonator with $t_{GaAs/AlGaAs}$ = 70.5 nm has lower frequencies of the fundamental, pitch, and fourth modes compared to the other two microresonators. The thinner and narrower microresonator cantilever has a smaller mass that shifts the modes to lower frequencies.

Decreasing the microresonator width from 12 µm to 6 µm increases the $R_{MAX}$ by ~ 9%, 22%, 7% for the microresonators with 102.8 nm, 82.1 nm, and 70.5 nm of GaAs/AlGaAs thicknesses, respectively, as shown in Fig. 5(b). For the same change in the microresonator width, the $f_{MAX}$ of the microresonators increases by ~ 20%, 47%, and 17%, respectively. Similarly, the lowest frequency, $f_L$, increases by increasing the microresonator widths from 6 µm to 12 µm as shown in Fig. 5(c). We observe that the decreasing the microresonator widths affects the microresonator parameters in the same way as increasing the microresonator lengths. However, the sensitivity to length variations is roughly 10 times larger than the sensitivity to width variations. Decreasing the microresonator widths from 12 µm to 6 µm increases the BWE by 1.8%, 1.4%, and 2.3% for the microresonators with 102.8 nm, 82.1 nm, and 70.5 nm of GaAs/AlGaAs thickness, respectively. Thus, for the fixed parameters of l = 55 µm, r = 32 µm, y = 3.75 µm in Fig.1, we have to choose the narrowest microresonator cantilever (w = 6 µm) to have more secure sub-SQL range with the desirable properties of low mode frequencies, high $R_{MAX}$, low $f_{MAX}$, low $f_L$, and relatively high BWE.

### 3.3. Size of Microresonator Mirror Pad

Figure 6 shows the effect of varying the radius of the microresonator mirror pad. Increasing the mirror pad radius from 12 µm to 42 µm decreases the frequency of the fundamental mode by ~ 60% for all three microresonators considered as shown in Fig. 6(a). The shift of the fundamental modes to lower frequencies is not as significant as the shift caused by changing the length of the microresonator cantilever but is larger than the shift caused by changing its width. The frequencies of pitch mode decrease initially and increase to roughly the same frequencies as the mirror pad radius increases to 42 µm. The frequency of the fourth mode increases by a factor of ~5, to a value of ~1MHz, at a mirror pad radius of 42 µm. Therefore, the increase in mirror radius is beneficial, not only because of the decrease in the frequency of the fundamental mode to below 1 kHz but also because of the increase in the frequency of the fourth and upper modes to above 200 kHz. However, increasing the radius of the mirror pad increases the mass of the mirror, and thereby decreases the ratio of QRPN to TN [30].

Increasing the radius of the mirror pad from 12 µm to 22 µm increases $R_{MAX}$ by a factor of ~5. $R_{MAX}$ then decreases to roughly the same value at a mirror pad radius of 42 µm, for all the microresonators as shown in Fig. 6(b). The $f_{MAX}$ of the microresonators increases by ~11% when changing the radius of the mirror pad from 12 µm to 42 µm. We observe that the sensitivity of $f_{MAX}$ to the radius of the mirror pad is similar to the sensitivity to the microresonator width but smaller than the sensitivity to the microresonator length. Increasing the radius of the mirror pad from 12 µm to 22 µm steeply decreases $f_L$ by a factor of ~2.8, which then increases by a factor of ~1.8 as the radius of the mirror pad increases from 22 µm to 42 µm [Fig. 6(c)]. Increasing the radius of the mirror pad from 12 µm to 32 µm increases the BWE by ~17%, ~35%, and ~40% for the microresonators with 102.8 nm, 82.1 nm, and 70.5 nm GaAs/AlGaAs thicknesses, respectively [Fig. 6(c)]. For the mirror pad with a 42 µm radius, the pitch mode frequency increases and decreases the BWE by ~12% for all three microresonators considered. Weobserve that the dependence of $R_{MAX}$, $f_{MAX}$, $f_L$, and BWE on the mirror pad radius is not monotonic due to different shifts in the frequency of the fundamental, pitch, and fourth modes as the mirror pad radius is varied. Thus, for the fixed parameters of l = 55 µm, w = 8 µm, y = 3.75 µm in Fig.1, the microresonator mirror pad with the 22 µm radius has the highest $R_{MAX}$, lowest $f_L$, relatively high BWE and $f_{MAX}$, and the desirable fundamental mode frequency below 1 kHz to suppress the heat generation from the laser beam.

Figure 7 shows the effect of varying the location of the laser beam, y in Fig. 1, on the microresonator mirror pad. Fig. 7(a) shows that there is no change in the frequency of mechanical modes when altering the laser beam location because the laser beam location is not a geometric parameter of the microresonator. The location of the laser beam, y in Fig. 1, affects the contribution of the fourth mode and changes its sharpness at the fixed frequency of the fourth mode [Fig. 7(a)].

The position-dependent coupling of the optical spring to the fourth mode is analogous to attaching a mechanical spring to different points on the microresonator. We observe that increasing the distance of the laser beam location from the center of the mirror pad sharpens the fourth mode and thereby widens the plateau of the sub-SQL region. Fig. 7(b) shows that shifting the laser beam location from 2.75 µm to 5.75 µm increases $R_{MAX}$ by ~50% to ~4.95 and decreases $f_{MAX}$ by ~48% to ~78 kHz for the microresonator with thinnest GaAs/AlGaAs pair. Fig. 7(c) shows that $f_L$ decreases by ~60% and the BWE increases by ~ 10% by shifting the laser beam location from 2.75 µm to 5.75 µm. Thus, for the fixed parameters of l = 55 µm, w = 8 µm, and r = 32 µm in Fig.1, the microresonator mirror pad with the laser beam location of y = 5.75 µm has more secure sub-SQL range with highest $R_{MAX}$, lowest $f_{MAX}$, lowest $f_L$, and highest BWE.

## 4. Summary and Conclusion

We have designed new microresonators using the micro-genetic algorithm and a finite element method model and studied the effect of changing the geometry of the microresonators on the sub-SQL region. We have illustrated the results of our analysis by calculating the spectrum of TN and SQL through summing the contribution of each mechanical mode in quadrature for the microresonators. We find that thinner and longer microresonators have lower frequencies of the fundamental, pitch, and fourth modes, and that the sensitivity of the mode frequencies to the microresonator length is more significant than the sensitivity to its width and mirror radius. The proposed microresonators outperform the previous design in the sub-SQL region, demonstrating lower TN under SQL, a broader sub-SQL region, and a sub-SQL region at lower frequencies. We find that the maximum ratio of TN to SQL ($R_{MAX}$) is increased, its frequency ($f_{MAX}$) is decreased, and the sub-SQL range (BWE) is increased by increasing the length of the microresonator cantilever, increasing the radius of the mirror pad, decreasing the width of the microresonator cantilever, and shifting the laser beam location from the mirror center. The larger $R_{MAX}$ and smaller $f_{MAX}$ contribute to having a secure range of sub-SQL regime and suppress the heat generation from the laser beam. The larger BWE extended to higher frequencies shows a promising microresonator for designing an unstable filter cavity that operates beyond the frequency limit of a few hundred kHz in the current experimental setup. Increasing the ratio of SQL to TN and tuning its frequency range serves as a testbed for not only QRPN measurements but also quantum nondemolition (QND) measurements and macroscopic quantum measurements. This approach promotes the possibility of testing reduction techniques to ultimately obtain a sensitivity below the SQL. The broadband, off-resonance sub-SQL region opens a new regime of precision measurement that is relevant for many practical sensing applications, including large scale interferometric gravitational wave detection.

**Figures**

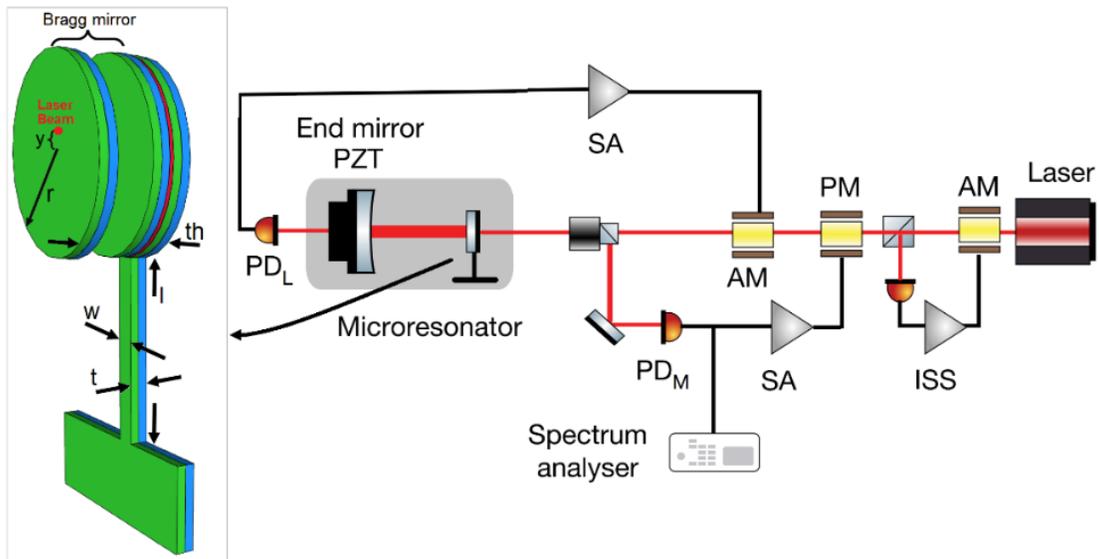

Figure 1. Experimental arrangement of our optomechanical system. The optomechanical cavity consists of a macroscopic end mirror and a microresonator [30]. The mirror pad consists of multiple layers of GaAs/Al$_{0.92}$Ga$_{0.08}$ quarter-wave optical pairs in a Bragg mirror, an InGaP buffer layer, and a GaAs/Al$_{0.92}$Ga$_{0.08}$As pair that also holds the cantilever microresonator. The geometric parameters of the microresonator are the mirror pad radius (r), its thickness (th), the distance of the laser beam spot from the mirror pad center (y), and the microresonator cantilever length (l), width (w), and thickness (t).

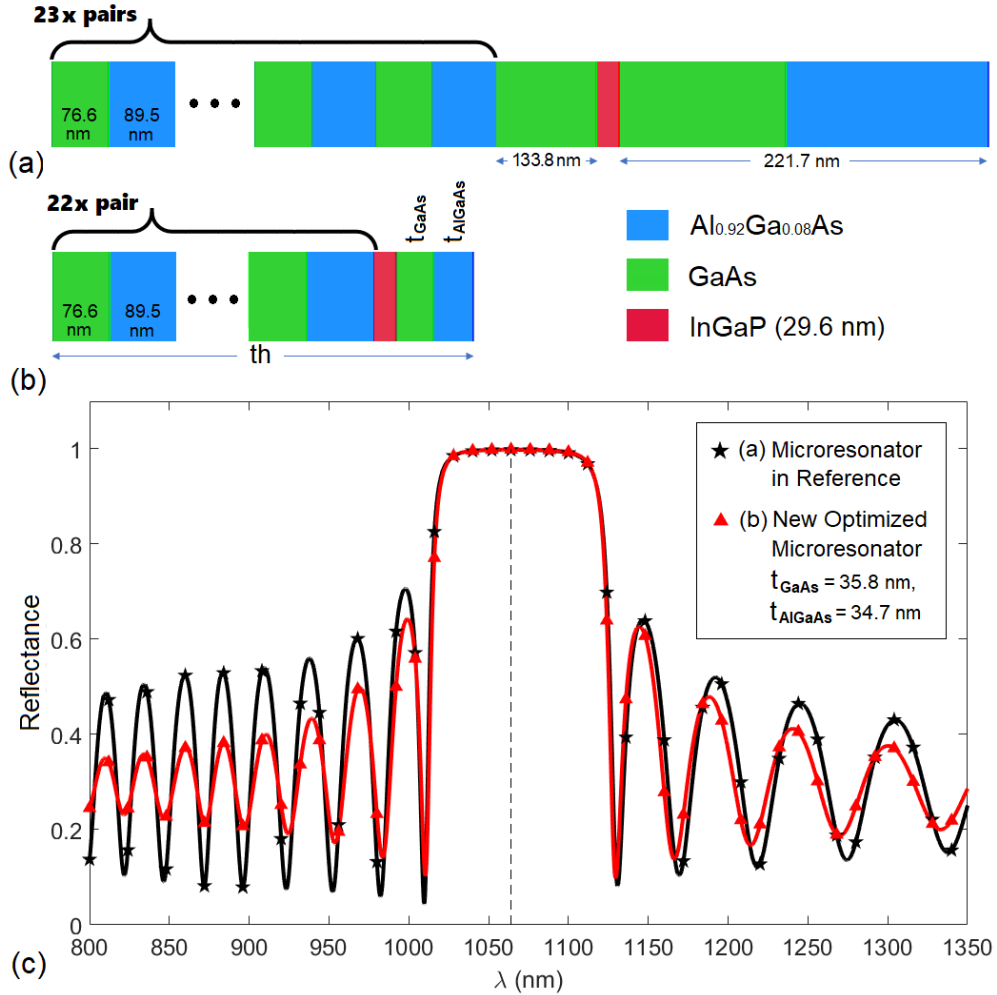

Figure 2. Cross-sectional schematic of (a) the microresonator mirror pad design in ref. [27] and (b) microresonator mirror pads designed using our optimization algorithm [41]. All the structures are designed for 99.975% reflectance at the center wavelength of 1078 nm. The aperiodic multilayer structures are composed of alternating layers of GaAs (76.6 nm), $Al_{0.92}Ga_{0.08}As$ (89.5 nm), and a lattice-matched InGaP etch stop layer (29.6 nm). (c) The reflectance as a function of wavelength in normal direction for the microresonator mirror pad structures in (a) and (b). The optimized thicknesses of the GaAs and $Al_{0.92}Ga_{0.08}As$ layers are 35.8 nm and 34.7 nm, respectively.

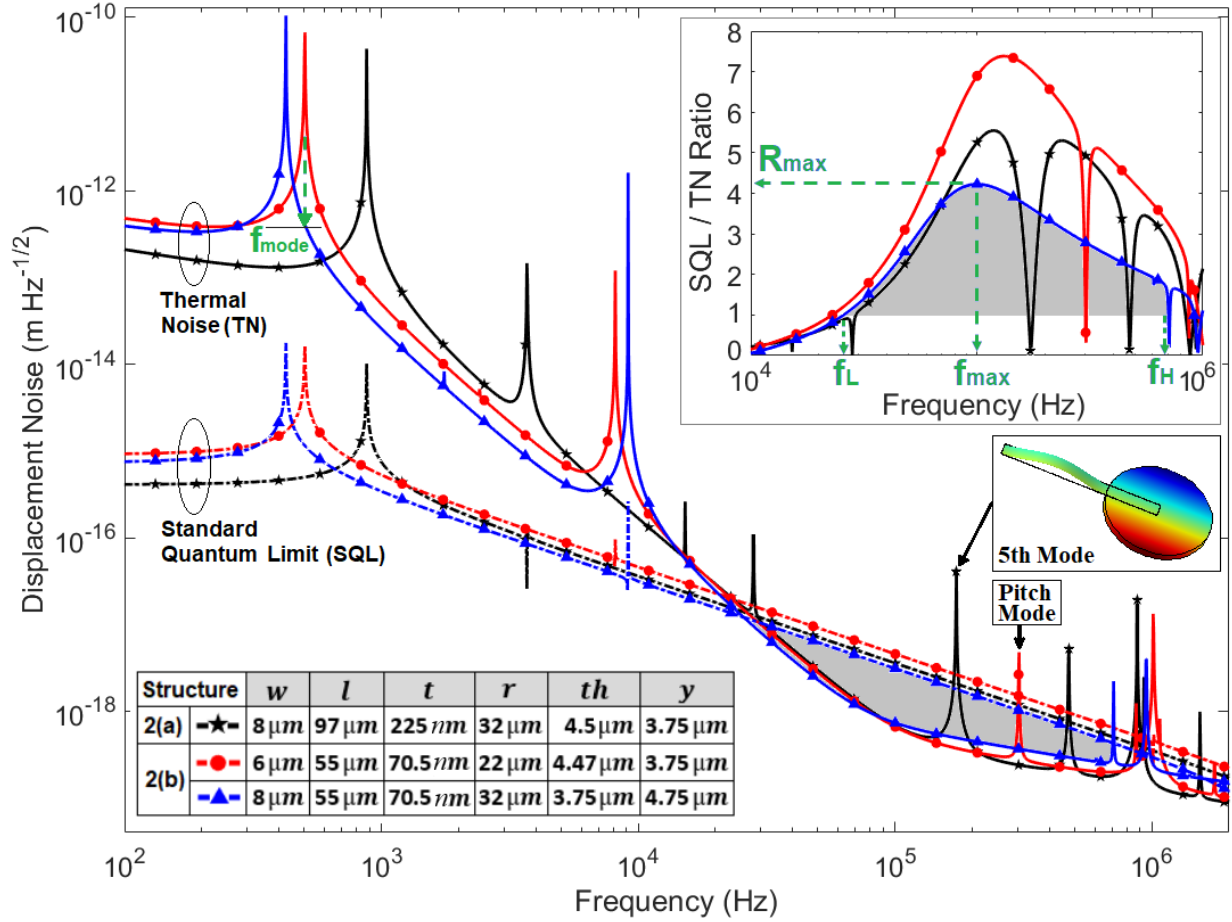

Figure 3. Thermal noise and standard quantum limit as a function of frequency for the microresonators in Figs. 2(a) and 2(b) with different geometries. The inset shows the SQL/TN ratio for the microresonators as well as the definition of the parameters used to study the effects of changing the geometry of the mirror pad and cantilever microresonator on the TN and SQL. The inset also shows the fifth mode of the microresonator in Fig.2(b) in the BWE region. These parameters include the maximum ratio, $R_{MAX}$, its frequency, $f_{MAX}$, the lowest frequency, $f_L$, and highest frequency, $f_H$, at which the TN equals to SQL. The two frequencies, $f_L$ and $f_H$, are used to calculate the frequency bandwidth of TN and SQL equality using $BWE = \log(f_H) - \log(f_L)$. The geometric parameters of the microresonators are shown in the inset table.

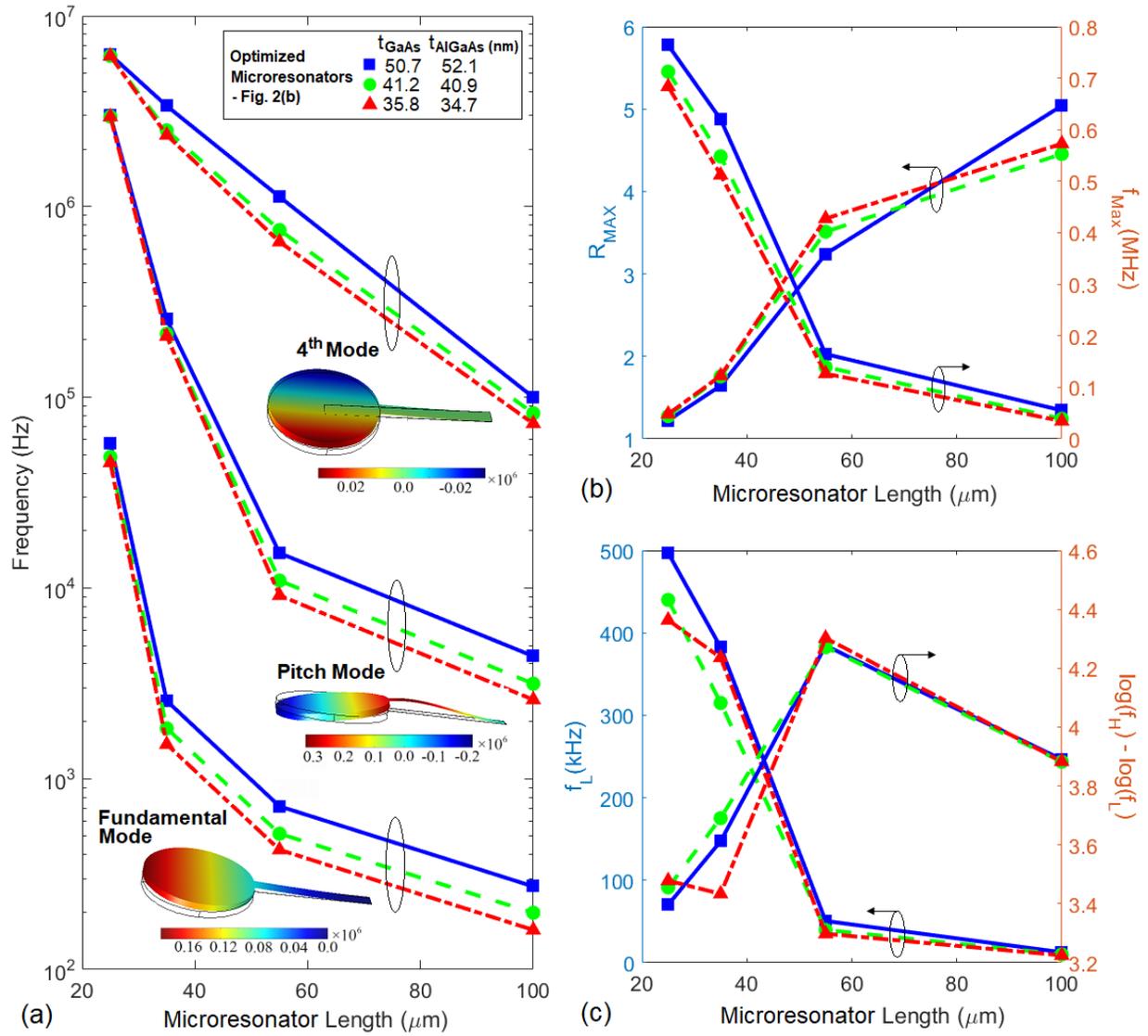

Figure 4. (a) Resonance frequencies of the fundamental, pitch, and fourth modes, (b) $R_{MAX}$ and $f_{MAX}$, (c) $f_L$ and BWE as a function of the microresonator cantilever length for the three microresonators optimized by the genetic algorithm. Results are shown for the fixed parameters of w = 8 μm, r = 32 μm, y = 3.75 μm. The microresonator mechanical modes of interest are also depicted in the figure.

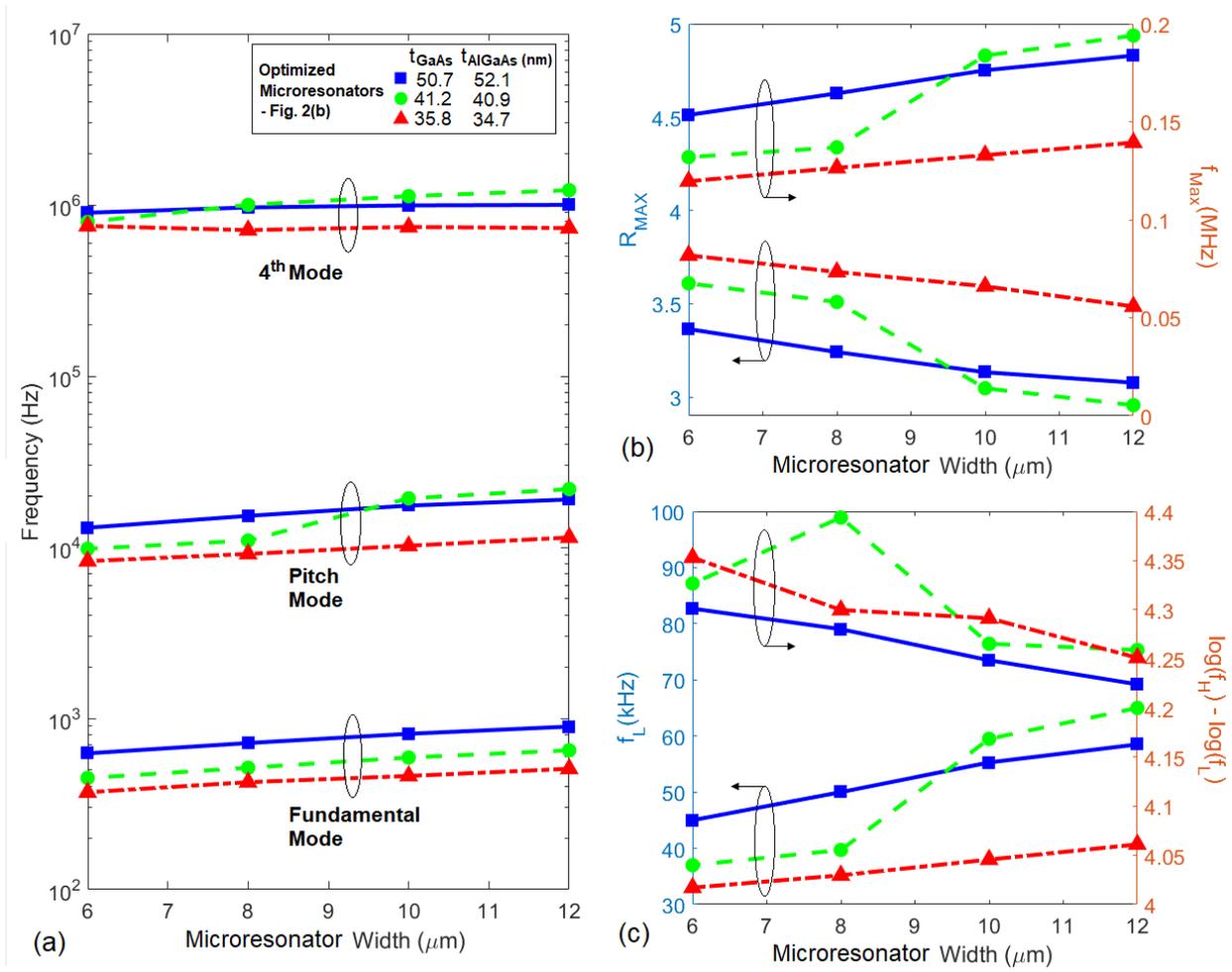

Figure 5. (a) Resonance frequency of the fundamental, pitch, and fourth modes, (b) $R_{MAX}$ and $f_{MAX}$, (c) $f_L$ and BWE as a function of the microresonator width for the three microresonators optimized by genetic algorithm. Results are shown for the fixed parameters of l = 55 µm, r = 32 µm, y = 3.75 µm.

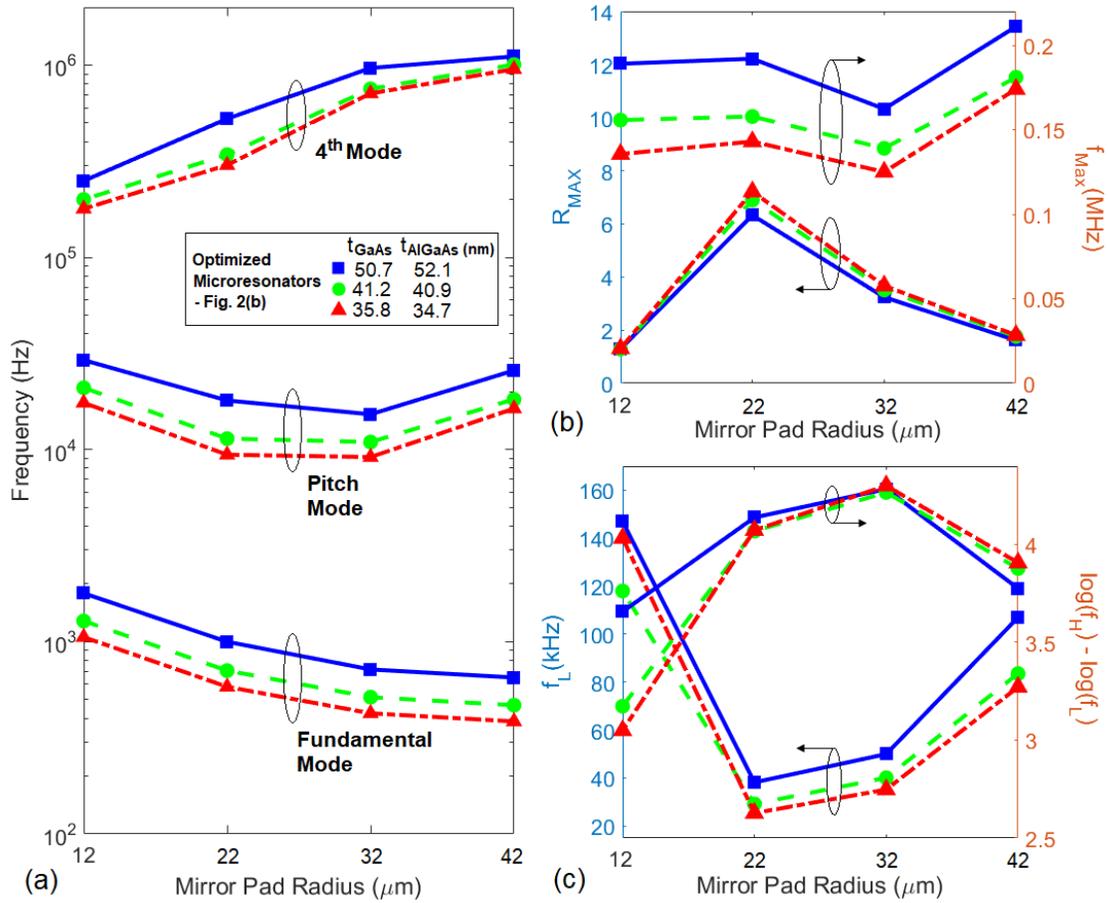

Figure 6. (a) Resonance frequency of the fundamental, pitch, and fourth modes, (b) $R_{MAX}$ and $f_{MAX}$, (c) $f_L$ and BWE as a function of the mirror pad radius for the three microresonators optimized by genetic algorithm. Results are shown for the fixed parameters of l = 55 μm, w = 8 μm, y = 3.75 μm.

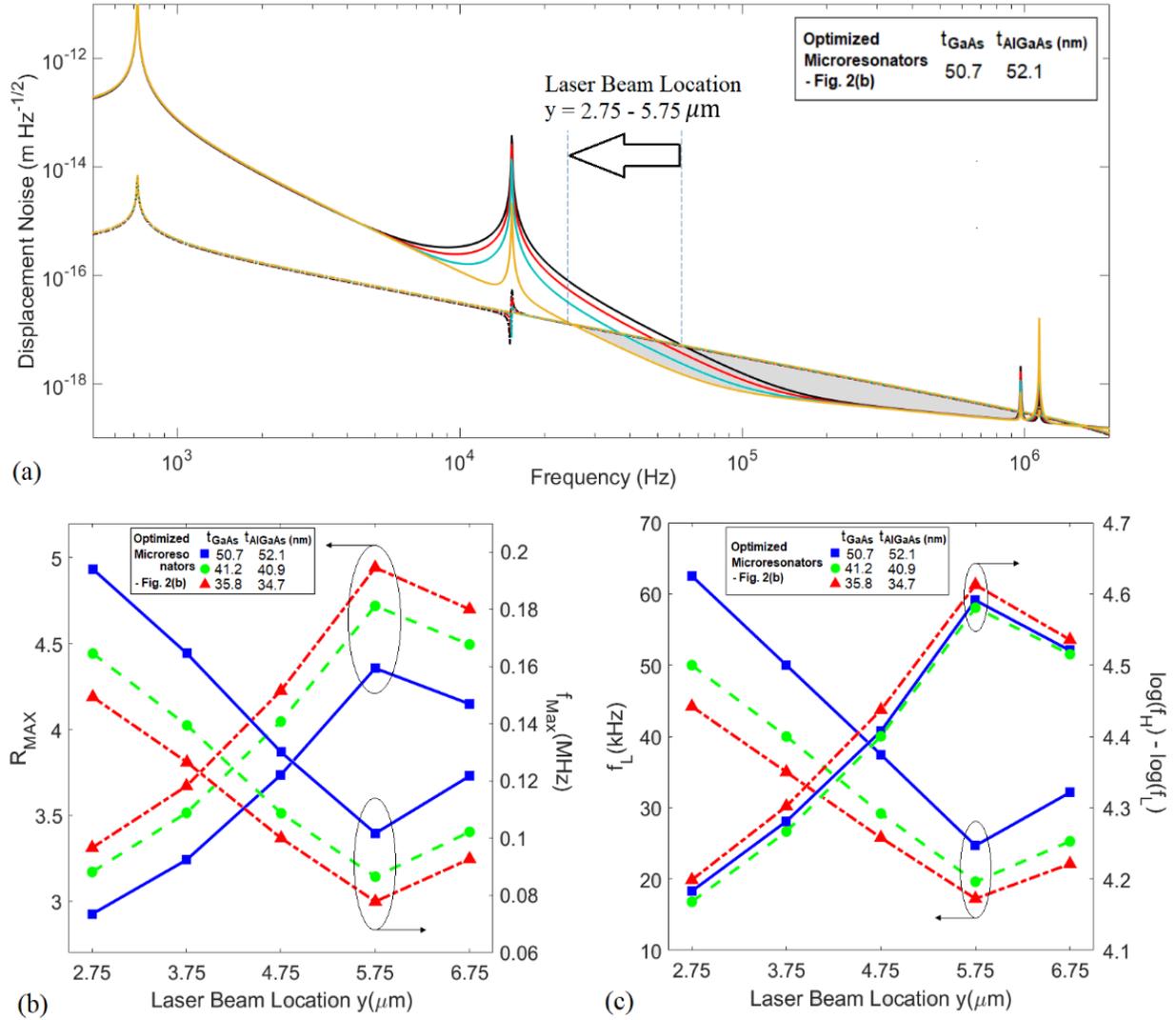

Figure 7. (a) Thermal noise and standard quantum limit as a function of frequency for the optimized microresonator in Fig. 2(b) with $t_{GaAs}$ = 50.7 nm and $t_{AlGaAs}$ = 52.1 nm. Moving the location of the laser beam away from the center of the mirror pad, from y=2.75 µm to 5.75 µm in Fig. 1, sharpens the fourth mode at the fixed frequency of this mode. (b) $R_{MAX}$ and $f_{MAX}$, (c) $f_L$ and BWE as a function of the laser beam location for the three microresonators optimized by genetic algorithm. Results are shown for the fixed parameters of l = 55 µm, w = 8 µm, and r = 32 µm.